\documentclass{aa}  
\usepackage{graphicx}
\usepackage{txfonts}
\usepackage{geometry}
\usepackage{placeins}
\usepackage{lscape}
\usepackage{hyperref}
\usepackage{xcolor}
\usepackage{ulem}

\hypersetup{
    colorlinks=true,
    linkcolor=blue,
    filecolor=magenta,      
    urlcolor=cyan,
    citecolor=blue,
    pdftitle={Overleaf Example},
    pdfpagemode=FullScreen,
    }

\makeatletter
\renewcommand*\aa@pageof{, page \thepage{} of \pageref*{LastPage}}
\makeatother

\begin{document}

    \title{Diversity of morphology of type II spicules in MURaM-ChE simulations}

    \author{Sanghita Chandra\inst{1},          
          Robert Cameron\inst{1},
          Damien Przybylski\inst{1}, 
          \and 
          Sami K. Solanki\inst{1,2}
          }

   \institute{Max Planck Institute for Solar System Research,
              Justus von Liebig Weg, 37077 G\"ottingen, Germany\\
              \email{chandra@mps.mpg.de}
              \and
              School of Space Research, Kyung Hee University, Yongin, Gyeonggi 17104, Republic of Korea
             }
    
   \date{Received: 04 Ferbruary 2026 / Accepted: 25 May 2026}

\abstract
{Spicules are ubiquitous, small-scale features in the solar atmosphere, exhibiting a jet-like appearance most clearly identified by their apparent motion in off-limb observations. While they are often interpreted as narrow, thread-like structures, their true three-dimensional (3D) structure remains unknown. }
{We aim to uncover the 3D morphology and dynamics of fast-evolving spicules (type II) using a MURaM-ChE simulation.}
{We use a H$\alpha$ proxy that has been developed using non-equilibrium (NE) hydrogen populations in MURaM-ChE. The proxy, modelled as an escape probability, is synthesised to isolate on-disc as well as off-limb H$\alpha$ wing features. The 3D structure of these features is investigated using the 3D information on opacity in H$\alpha$.}
{We identify type II spicules, with unique 3D morphologies: the dominant being thread-like and slab-like. The appearance of spicules as slabs or threads is a function of time and Doppler velocity. The spicules extending above the spicule-forest (2\;Mm\;-\;3\;Mm above the surface) tend to be located at quasi-separatrix layers (QSLs). We find that the spatially resolved contributions to the opacity of spicules are often similar for spicules synthesised in the horizontal direction, and their on-disk rapid blueshifted excursion (RBE) synthesised in the vertical direction at the same Doppler velocity of 37\;km/s. This confirms that RBEs are indeed the on-disc counterparts of spicules. Furthermore, our analysis indicates that cross-field motions can significantly contribute to spicule dynamics.}
{Spicules exhibit a range of morphologies, including both slab-like and thread-like structures. Their observed appearance depends strongly on line-of-sight projection and Doppler sampling. Spicules are preferentially located at QSLs, highlighting the role of magnetic topology in driving spicular dynamics.
}


\titlerunning{Diversity of morphology of type II spicules}
\authorrunning{Chandra et al.}

   \maketitle
%

\section{Introduction}

Spicules are highly dynamic, jet-like structures that originate in the solar chromosphere. They exhibit complex motions, including torsional and transverse components, which make them challenging to track using intensity images alone. A subclass known as type II spicules \citep{de_Pontieu_2007a} exhibits rapid apparent upward motion when observed at the solar limb, often extending into the upper atmosphere. Many of these features, which appear thin and hair-like, fade from the passband in which they are initially detected, with some reappearing in transition-region diagnostics \citep{Pereira_2014}. However, this behaviour is not universal. A detailed understanding of their three-dimensional (3D) geometry and dynamics is therefore essential for interpreting their observational signatures, including the rapid fading seen in chromospheric passbands. Such insight would place stronger constraints on the contribution of spicules to the mass and energy budgets of the chromosphere and corona, and on their potential role in coronal heating.

Spicules were first formally described by \citet{secchi_1877}. Subsequently, the reviews by \citet{Beckers_1968, Beckers_1972} provided a comprehensive description of their properties based on the early observations with hydrogen, helium, and calcium spectral lines, characterising their abundance, lifetimes, and morphology. A key challenge in studying such small-scale chromospheric structures arises from the intrinsic limitations of remote-sensing observations. The superposition of multiple spicules at the limb complicates the interpretation of their line-of-sight (LoS) motions, while plane-of-sky observations alone do not reveal the true plasma velocity. These limitations hinder our ability to reconstruct their full 3D structure and dynamics. Traditionally, spicules have been regarded as thin, tubular features aligned with the magnetic field. However, \citet{Judge_2011, Judge_2012, Lipartito_2014} proposed an alternative interpretation, suggesting that spicules may instead represent tangential discontinuities that \citet{Parker_1994} associated with warped magnetic sheets. This idea was then challenged by \citet{Pereira_2016}, who argued that spicules are predominantly mass motions and that their complex swaying and apparent disappearance can be explained without invoking a sheet-like geometry. While the latter study presented observational evidence of swaying spicules supporting their interpretation, observational limitations make it difficult to determine the exact 3D morphology and motion of spicules.

Despite the challenges in interpreting spicules in 3D, studies have been conducted to understand spicules on a statistical basis. Some of the statistical properties inferred from early observational studies include typical lifetimes (3-10 min), maximum heights (5-10\;Mm), and velocities (10-40\;km/s). These relatively long-lived spicules, whose on-disc counterparts are identified as dynamic fibrils, are now known as type I spicules. With the advent of high-resolution observations from the \textit{Hinode} mission and the Swedish 1-m Solar Telescope (SST), a rapidly evolving subclass of spicules was discovered, that were found to rapidly fade out of the Ca II H passband at their maximum heights. This led to the classification of spicules into types I and II \citep{de_Pontieu_2007a}. However, type II spicules often display similar characteristics as the type I spicules when observed at low resolutions \citep{Pereira_2013}. Type I spicules, or dynamic fibrils are the canonical, less-dynamic spicules which were mostly found near active regions. Type II spicules -- the focus of this work --  are short-lived, jet-like features with lifetimes of 10-150 seconds, and apparent upward velocities of 30-150\;km/s. They are currently of greater interest because they appear to rapidly fade into the upper solar atmosphere. A heating signature in Si IV diagnostics associated with a fraction of these spicules suggests that they may be heated to transition region temperatures \citep{Pereira_2014, Rouppe_van_der_Voort_2015}. The on-disc counterparts of type II spicules are thought to be the rapid blueshifted excursions (RBEs), with the redshifted counterparts identified as rapid redshifted excursions (RREs) \citep{Rouppe_van_der_Voort_2009, Sekse_2012,Sekse_2013a, Danilovic_2023}. RREs are believed to be produced when the line-of-sight component of the often slanted upflows are redshifted. The rapid excursions are identified in spectral lines such as H$\alpha$, Ca II H, etc. Although RBEs are deemed to be the on-disc counterparts of type II spicules based on the similarities in their morphology and timescales of evolution, there is no confirmation that this is indeed the case. We are restricted by the lack of stereoscopic observations or numerical models which accurately reproduce all the observed properties of spicules. 


    \begin{figure*}
   \centering
   \includegraphics[width=\textwidth]{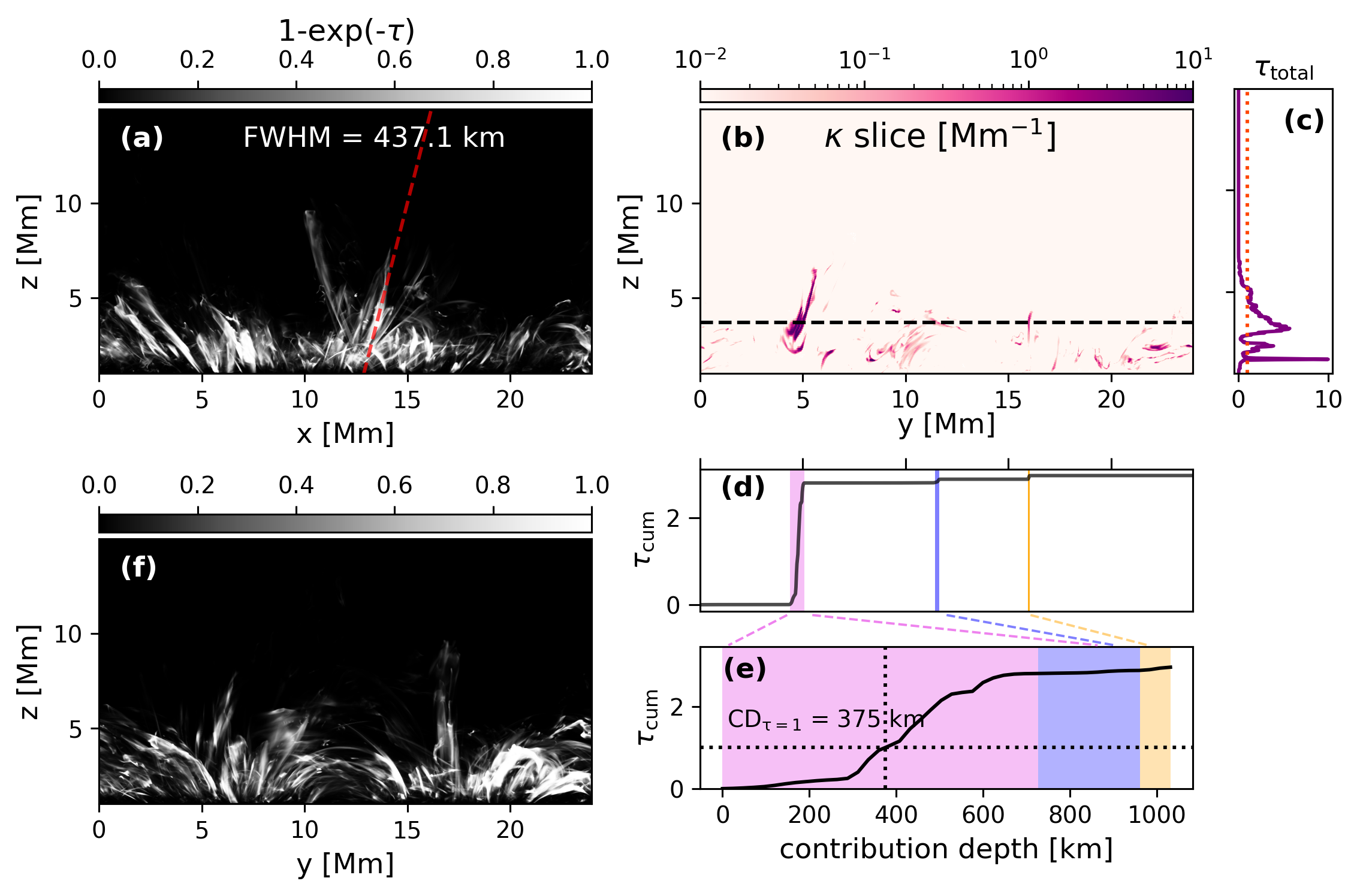}
   \caption{Overview of a singly-threaded spicule. In panel (a), the H$\alpha$ proxy ($\mathsf{v}_\mathrm{D}$ = 37\;km/s) image shows the spicule; its width is defined by the full width at half maximum (FWHM; see text). The dashed red line marks the line-of-sight slice shown in panel (b). In panel (b), a slice of the opacity ($\kappa$) along the line of sight illustrates the plasma contributing to the spicule. The total optical depth ($\tau_{\mathrm{total}}$) as a function of height is shown in panel (c), with the $\tau = 1$ level indicated by the dotted red line. The dashed black line in panel (b) marks the cut through the $\kappa$ slice used to compute the cumulative optical depth ($\tau_{\mathrm{cum}}$) shown in panel (d). In panel (d), $\tau_{\mathrm{cum}}$ along the cut indicated in panel (b) is shown, with the observer at $y\;=\;0\;\mathrm{Mm}$ looking towards $y\;=\;24\;\mathrm{Mm}$. In panel (e), the same quantity as in panel (d) is shown, but with an opacity threshold of $\kappa > 0.1\;\mathrm{Mm^{-1}}$  applied to isolate the dominant contribution to the spicule. The corresponding regions are shaded in various colours in panels (d) and (e). The thickness of the spicule along the line of sight is indicated as the contribution depth ($\mathrm{CD_{\tau = 1}}$) in panel (e). The dotted black horizontal and vertical lines mark the $\tau = 1$ level and the corresponding contribution depth respectively. For comparison with the opacity contribution, we also show the orthogonal view of the spicule in panel (f), on the bottom left. Panels (a) and (f) share the same colour bar and are orthogonal views to each other.}
              \label{thread-535000}%
    \end{figure*}

    \begin{figure*}
   \centering
   \includegraphics[width=\textwidth]{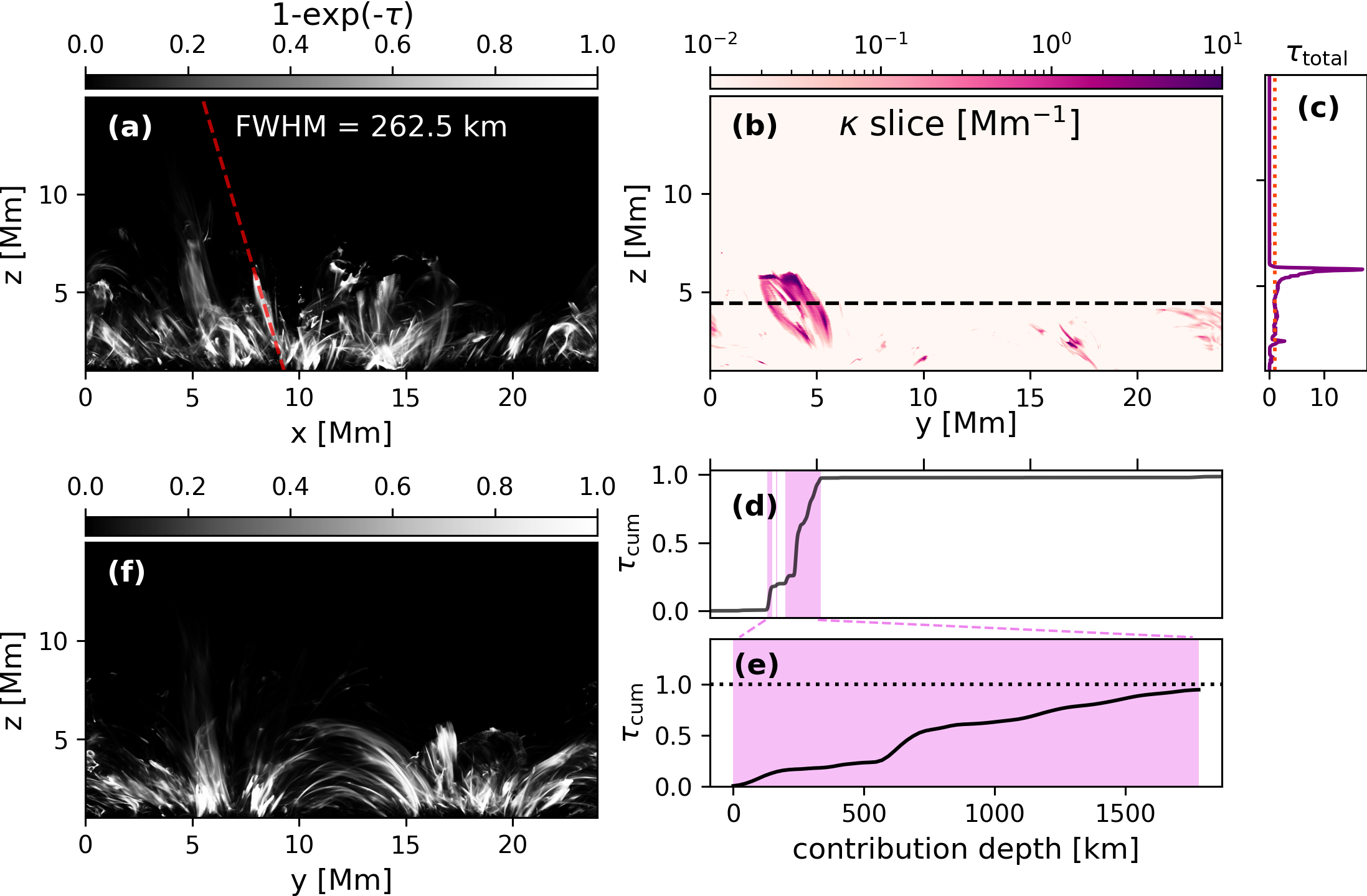}
   \caption{Overview of a slab-like spicule (spicule A; see text) in the same format as Fig. \ref{thread-535000}. The contribution depth of the spicule ($\mathrm{CD_{\tau = 1}}$) is given by the entire extent shown in panel (e), as the optical depth does not reach $\tau = 1$ at the height indicated in panel (b) by the dashed horizontal black line.}
              \label{sheet-510000}%
    \end{figure*}


While statistical studies and classifications have established the basic properties of spicules, their underlying dynamics and driving mechanisms remain debated. The dynamics of spicules reflect the underlying driving mechanisms that generate them. Observations reveal a combination of field-aligned, transverse, and torsional motions, whose interplay provides clues to the dominant forces at work. Early numerical studies aimed at understanding the driving mechanisms of spicules are reviewed in \citet{Sterling_2000}. More recent efforts, combining imaging spectroscopy \citep{Sekse_2013b} with simplified numerical models \citep[e.g.][]{Sharma_2017}, have described the interplay of different kinds of spicule motions and possible associated wave driving. Generally, the proposed drivers for spicules include magnetic reconnection \citep{Gonzalez_2017, Samanta_2019}, and magnetohydrodynamic (MHD) waves \citep{Zaqarashivli_2009}. Multi-fluid effects such as ambipolar diffusion and ion–neutral interactions have also been proposed to mediate the build-up and release of magnetic tension to drive spicules \citep{Sykora_2017, Sykora_2020}. Previous studies have shown that spicules frequently exhibit strong transverse motions \citep{Antolin_2018}, which could be connected to waves and instabilities. On the contrary, the jet-like or upward and occasional downward displacements, when observed off-limb, have motivated models describing them as field-aligned flows within relatively static cylindrical flux tubes \citep{Sterling_1993, Kudoh_1999}. It remains unclear what forces and flows drive spicule dynamics.

Another long-standing question concerns the apparent sudden disappearance and reappearance of type II spicules. One explanation is heating to coronal temperatures, leading to mass and energy transport into the corona. Other explanations build on visibility effects. Some studies invoke line-of-sight effects due to integrating through a warped magnetic sheet \citep{Judge_2011}, others explain it with strong transverse motions, similar to the cause of disappearance from a passband \citep{Pereira_2016}. Yet other simulations suggest that line-of-sight effects from integrating through spinning jets can give the appearance of spicules \citep{Iijima_2017, dey_2024}. It is crucial to infer the 3D morphology and motion of spicules to understand their fundamental nature. 

In this work, we address several of the open questions outlined above by studying the 3D morphology of spicules on a statistical basis, while examining the dynamics of a few representative examples. We also look into the magnetic environment surrounding spicules. We investigate type II spicules using a 3D radiative-MHD simulation of an enhanced network region, performed with the chromospheric extension of MURaM (MURaM-ChE; \citealt{Przybylski_2022}). The model atmosphere is described in Sect. \ref{model_atmosphere}. To identify spicules and analyse their 3D morphology, we employ a synthetic H$\alpha$ observable at the limb (Sect. \ref{Proxy}). We identify spicular structures with both slab-like and thread-like morphologies, and report on the morphology of 58 spicules (Sect. \ref{feature-of-interest}). We examine two representative spicules in more detail in Sect.~\ref{dynamics_spicules}, and compare their appearance in the plane-of-sky and in the line-of-sight. In Sect. \ref{discussion}, we discuss the implications of our results, and in Sect. \ref{conclusion} we summarise our conclusions.

\section{Enhanced Network Region: Model Setup}\label{model_atmosphere}

The simulation analysed in this work was carried out with MURaM-ChE \citep{Przybylski_2022}, the chromospheric extension of the MURaM radiative-MHD code \citep{Voegler_2005, Rempel_2017}. This version of MURaM includes a time-dependent treatment of hydrogen level populations via solution of the non-equilibrium (NE) rate equations, a crucial ingredient for correctly modelling H$\alpha$ line formation \citep{Leenaarts_2007}. This treatment is important for our calculations, where the NE hydrogen populations form the backbone of the H$\alpha$ proxy. We identify naturally occurring spicules with the H$\alpha$ proxy in the simulation. That is, no driver was introduced into the simulation, which was in no way adapted or tweaked to produce spicules. These spicules arise naturally from the interaction of turbulent convection with the magnetic field that reaches into the chromosphere and corona.  
 
The model represents a solar enhanced network region, with a bipolar magnetic configuration superimposed on a 2x2 tiling of the small-scale dynamo setup presented in \citet{Przybylski_2025}. The average flux imbalance in the box is $-$0.18\;G. The setup shares similarities with the publicly available Bifrost enhanced network model \citep{Carlsson_2016}. The horizontal resolution in the MURaM enhanced network model is two times higher than that of Bifrost. The magnetic field strength $|{\bf{B}}|$ at the average height of the $\tau_{500}=1$ surface ($z=0$) is $103.7$~G, about twice that of the Bifrost enhanced network model. The MURaM-ChE simulation does not incorporate a generalised Ohm's law \citep{Braginskii_1965}, as in the 2D simulations of spicules performed by \citet{Sykora_2017_ApJ, Sykora_2017}. We note that the results shown here were obtained without including ambipolar diffusion for our study.

The computational domain spans $24\times24\times24$~Mm$^{3}$, extending from 7~Mm below the mean $\tau_{500}=1$ surface or photosphere to 17~Mm above, into the corona. The numerical resolution is uniform at 23.46~km horizontally and 20~km vertically, enabling us to resolve the photospheric and chromospheric fine structure. The side boundaries are periodic, while the lower boundary is open with a symmetric magnetic field condition \citep{Rempel_2014}. The upper boundary is open to outflows, closed to inflows and uses a potential-field extrapolation for the magnetic field. Further details of the enhanced network model can be found in \citet{Ondratschek_2024}.

For this study we use a time sequence of 206 snapshots, covering 1280~s of solar evolution with a cadence of roughly 6~s. The same simulation has previously been used for forward modelling of the Ca II 8542 \AA\ infrared line \citep{Ondratschek_2025}, and the Mg II h \& k lines \citep{Ondratschek_2024}. The simulation has also been employed in the study of RBEs and synthetic H$\alpha$ wing features \citep{Chandra_2025}, and the statistics of off-limb spicules in \citet{Chandra_2026}.

\section{The H$\alpha$ proxy}\label{Proxy}

We employ a H$\alpha$ proxy, synthesised using the NE hydrogen populations, temperature, and line-of-sight velocity computed with MURaM-ChE. This proxy offers a physically motivated and computationally efficient means of identifying transient H$\alpha$ features forming in the line wings of the synthesised spectrum. The approach was originally developed for studying on-disc phenomena \citep{Chandra_2025} and later adapted for the analysis of off-limb structures \citep{Chandra_2026}. The formulation and implementation of the proxy are described comprehensively in these works.

We model the line profile as a function of the Doppler velocity. This profile function, $\Phi$($\mathsf{v}_\mathrm{D}$), models the Doppler broadening based on the temperature and the velocity of the plasma. We then compute the absorption coefficient, $\kappa$\footnote{We denote the opacity $\kappa$, by the absorption coefficient (commonly written as $\chi$~[$\mathrm{Mm^{-1}}$]) which differs from the standard convention (see \citealp[]{Mihalas_1978}).} using the NE hydrogen populations in MURaM-ChE. 
We compute the optical depth, $\tau$ by integrating the opacity ($\kappa$) in the line of sight. For the on-disc view, the line of sight is the vertical (z) direction. For off-limb spicules the horizontal (x or y) directions serve as the line of sight. The proxy in absorption is defined as,
\begin{align}\label{proxy}
    & & P(s) = \mathrm{exp}(-\tau),
\end{align}

\noindent \textit{P(s)} is the probability of photon escape when a photon travels a distance s in the atmosphere. This is the formulation for studying absorption features with the H$\alpha$ proxy. The proxy at the limb is defined to be (\textit{1$-$P(s)}), which is aimed to model the scattered photons along the line of sight, in the H$\alpha$ spectral line. Further details of identifying off-limb spicules with the H$\alpha$ proxy is explained in \citet{Chandra_2026}.

\section{Spicule morphology}\label{feature-of-interest}

We analysed the morphology of 58 spicules in the simulation. To probe the morphology in 3D, we use the opacity ($\kappa$) data cube, extract a slice through the feature, and analyse the line-of-sight contributions in the opacity. Most of these spicules (54 out of 58) are type II spicules with linear trajectories. However, we report the morphology of all 58 spicules we detected in our earlier work \citep{Chandra_2026}. The analysis for identifying 3D spicule morphology is shown in Figs. \ref{thread-535000} and \ref{sheet-510000}, showing an example singly-threaded and slab-like spicule, respectively. In the following, we describe the method of spicule classification in terms of their morphology based on Fig. \ref{thread-535000}. The same description applies for Fig. \ref{sheet-510000} (and Figs. \ref{multithread}, \ref{sheet-thread}).

We first detected the spicules in the proxy at the limb (panel a, Fig. \ref{thread-535000}). The synthesis was done for a Doppler velocity ($\mathsf{v}_\mathrm{D}$) of 37~km/s. The width of the spicule is reported as the full width at half maximum (FWHM) of the Gaussian fit to the intensity profile perpendicular to the spicule axis (see \citealp[for more details]{Chandra_2026}). We took an opacity ($\kappa$) slice through the spicule in the line of sight, such that the slice is placed along the spicule axis. This is shown by the dashed red line in panel (a). In panel (b) the opacity slice shows the contribution along the line of sight. The total optical depth ($\tau_{\mathrm{total}}$) as a function of height is shown in panel (c), with the dotted red line marking the $\tau = 1$ level. This allows us to infer the dominant contribution to the spicule up to the point where it becomes opaque ($\tau = 1$). Next, we check the optical depth at a particular height marked by the dashed horizontal black line on panel (b). To quantify this, we also plot the cumulative optical depth ($\tau_{\mathrm{cum}}$) for an observer at $y=0\;\mathrm{Mm}$ as a function of the distance along the line of sight in panel (d). We then imposed a threshold of $\kappa > 0.1\; \mathrm{Mm^{-1}}$ to isolate the dominant contribution to the spicule. This is highlighted by the shading on panel (d) in various colours, denoting different spatial regions. The corresponding optical depth from these regions is then shown on panel (e), where we show the thickness of the spicule along the line of sight with a quantity called the contribution depth (CD). The thickness of the spicule itself ($\mathrm{CD_{\tau = 1}}$) is identified by the location where $\tau = 1$, marked by the dotted horizontal black line in panel (e). 
If the feature does not become optically thick, i.e. 
$\tau = 1$ is not reached, then we define 
$\mathrm{CD_{\tau = 1}}$ accounting only for regions where $\kappa > 0.1\; \mathrm{Mm^{-1}}$.

In Fig.~\ref{thread-535000}, we show the contribution depth for an observer at $y=0\;\mathrm{Mm}$ looking towards $y=24\;\mathrm{Mm}$ (+y axis). If the observer looks from $y=24\;\mathrm{Mm}$ towards $y=0\;\mathrm{Mm}$ ($-$y axis), $\mathrm{CD_{\tau = 1}}$ will in general be different except in those cases where the feature does not become optically thick (see Appendix, Fig.~\ref{sheet-thread}). Next, we compute the ratio of the thickness along the line of sight (or contribution depth) to the spicule width (FWHM). If this ratio exceeds 2.5, we call it a slab-like spicule. Otherwise it is a singly-threaded spicule. Sometimes, spicules can be composed of multiple threads aligned along the line of sight, which are spatially separated in the opacity ($\kappa$) slice (multi-threaded spicules). We also find a fourth category where the spicules appear either as slabs or as threads depending on the viewing direction (e.g. along +y axis or $-$y axis) in the $\kappa$ slice. The appearance of the spicules in an orthogonal off-limb view is shown in panel~(f) of Figs.~\ref{thread-535000} and \ref{sheet-510000}.


\begin{table}[ht!]
\caption{Statistics on the morphology of spicules.}\label{table:1}
\centering
\begin{tabular}{ |c|c| } 
 \hline
 Morphology & Fraction  \\

 \hline
 \hline
 Singly-threaded  & 44.8\% \\
 Slab-like & 29.3\% \\ 
 Multi-threaded & 5.2\% \\
 Slab/thread, depending on LoS& 20.7\%\\
 \hline
\end{tabular}\
\vspace{0.1cm}
\tablefoot{Population of spicules (out of a total of 58 spicules) with different categories of spicule morphology.}
\end{table}


This analysis was performed at a fixed position along the spectral line, corresponding to a Doppler velocity of 37 km/s, using a single snapshot for each spicule taken at the moment it reached peak intensity. The height at which the optical depth was calculated (dashed black line in Fig.~\ref{thread-535000}) was chosen visually such that it passes through the maximum extent of our spicule of interest, while minimising adjoining overlapping structures. Within these constraints, our analysis revealed four spicule categories: (a) singly-threaded, (b) slab-like, (c) multi-threaded, or (d) a complex morphology which appears as either slab-like, or thread-like, depending on the direction of viewing. The fraction of spicules with each type of morphology is shown in Table~\ref{table:1}. We found the spicules to be dominantly singly-threaded, followed by the slab-like spicules. However, it is to be noted that during their evolution spicules can change between being thread-like and slab-like. The distribution of spicule widths (FWHM) and contribution depths ($\mathrm{CD_{\tau=1}}$) for all 58 spicules are shown in Fig.~\ref{distribution}.


    \begin{figure*}
   \centering
   \includegraphics[width=\textwidth]{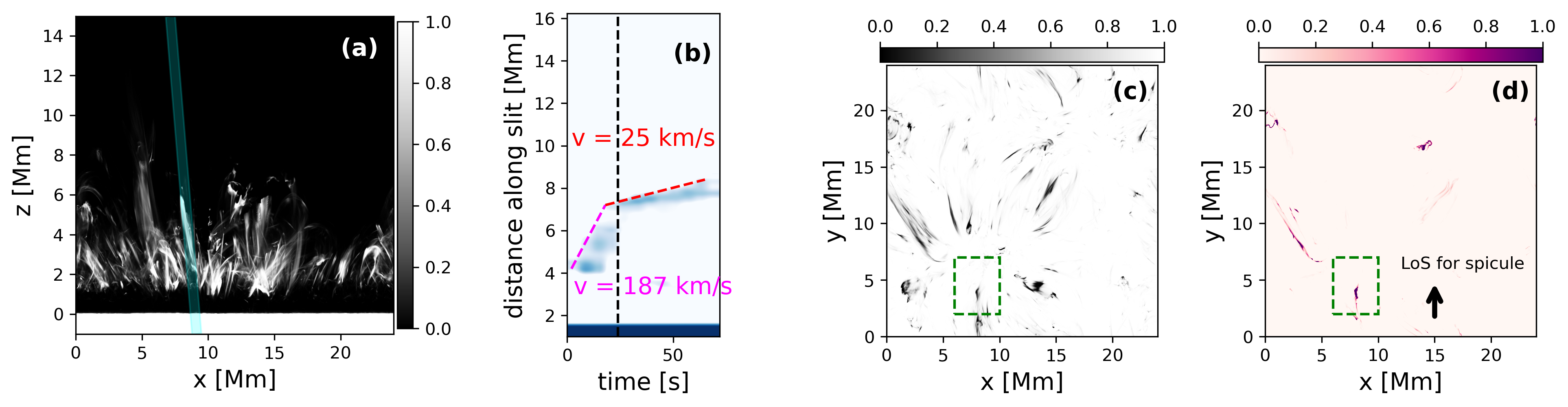}
   \caption{Overview of the evolution of the slab-like spicule A. Panel (a) shows a feature resembling a spicule identified with the H$\alpha$ proxy at the limb at a Doppler velocity ($\mathsf{v}_\mathrm{D}$) of 37 km/s. The artificial slit placed over the spicule is shown in cyan. Panel (b) depicts the time-distance evolution of this spicule along the artificial slit, with the apparent speeds marked by the dashed lines in red and magenta. Panel (c) shows the RBE counterpart of the spicule-like feature obtained at the same $\mathsf{v}_\mathrm{D}$ in the on-disc synthesis. Panel (d) shows an opacity ($\kappa$) slice at a constant height of 5.3\;Mm. The dashed green box in panels (c) and (d) shows the location of the spicule of interest. The evolution of the spicule and its RBE counterpart can be found in the movie \href{https://owncloud.gwdg.de/index.php/s/KhLwZhfeFHzIBtR}{online}.}
              \label{overview_F1}%
    \end{figure*}

    \begin{figure*}
   \centering
   \includegraphics[width=\textwidth]{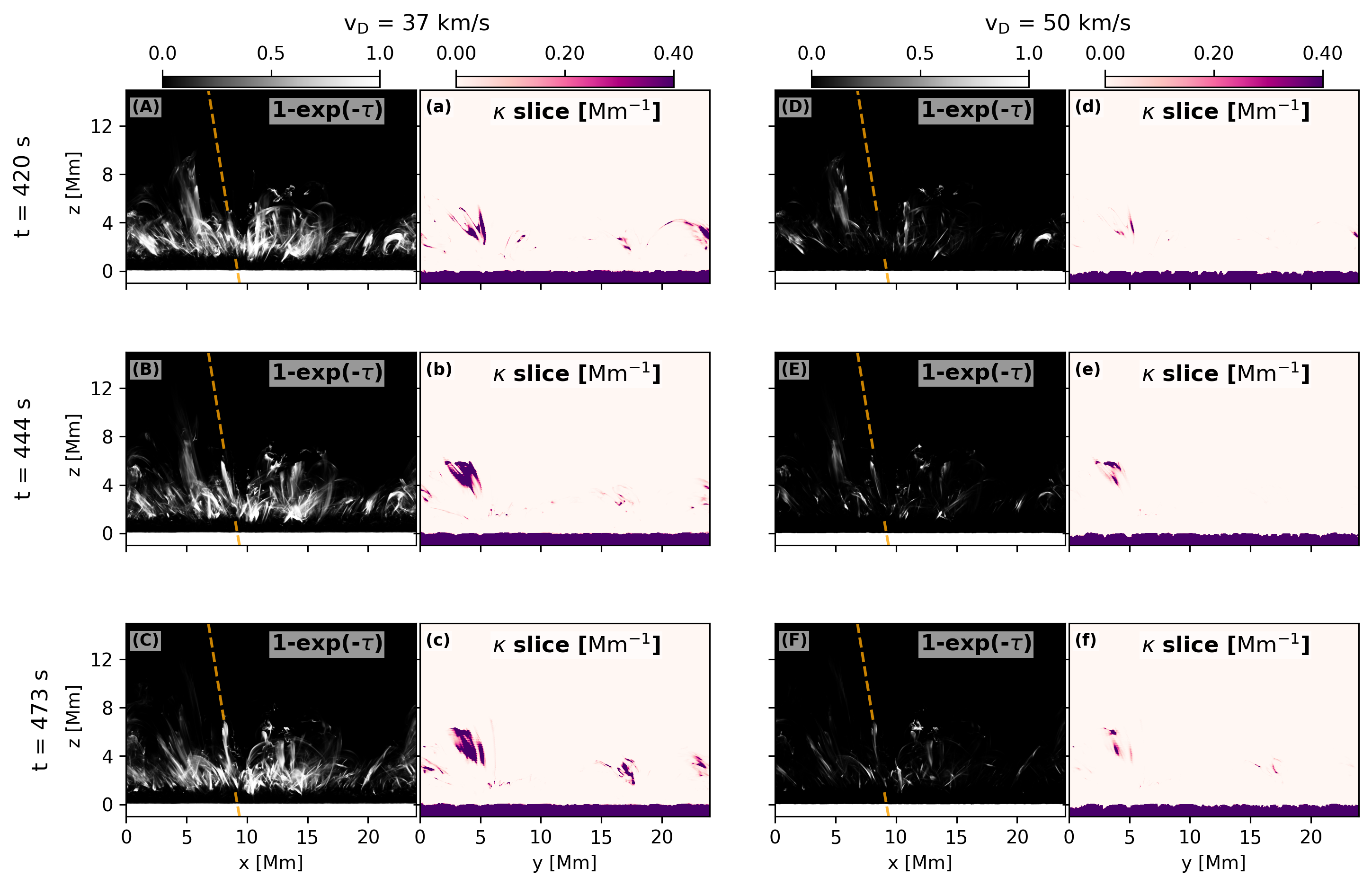}
   \caption{Evolution of spicule A at different Doppler velocities. The leftmost pair of columns corresponds to $\mathsf{v}_\mathrm{D}$ = 37 km/s, while the rightmost pair of columns correspond to $\mathsf{v}_\mathrm{D}$ = 50 km/s. The dashed orange line in the panels with uppercase letters marks the slice through the spicule. The line-of-sight opacity contribution to the spicule ($\kappa$ slice) is shown in the corresponding panels with lowercase letters. 
   }
              \label{F1_various_vD}%
    \end{figure*}

    \begin{figure*}
   \centering
   \includegraphics[width=\textwidth]{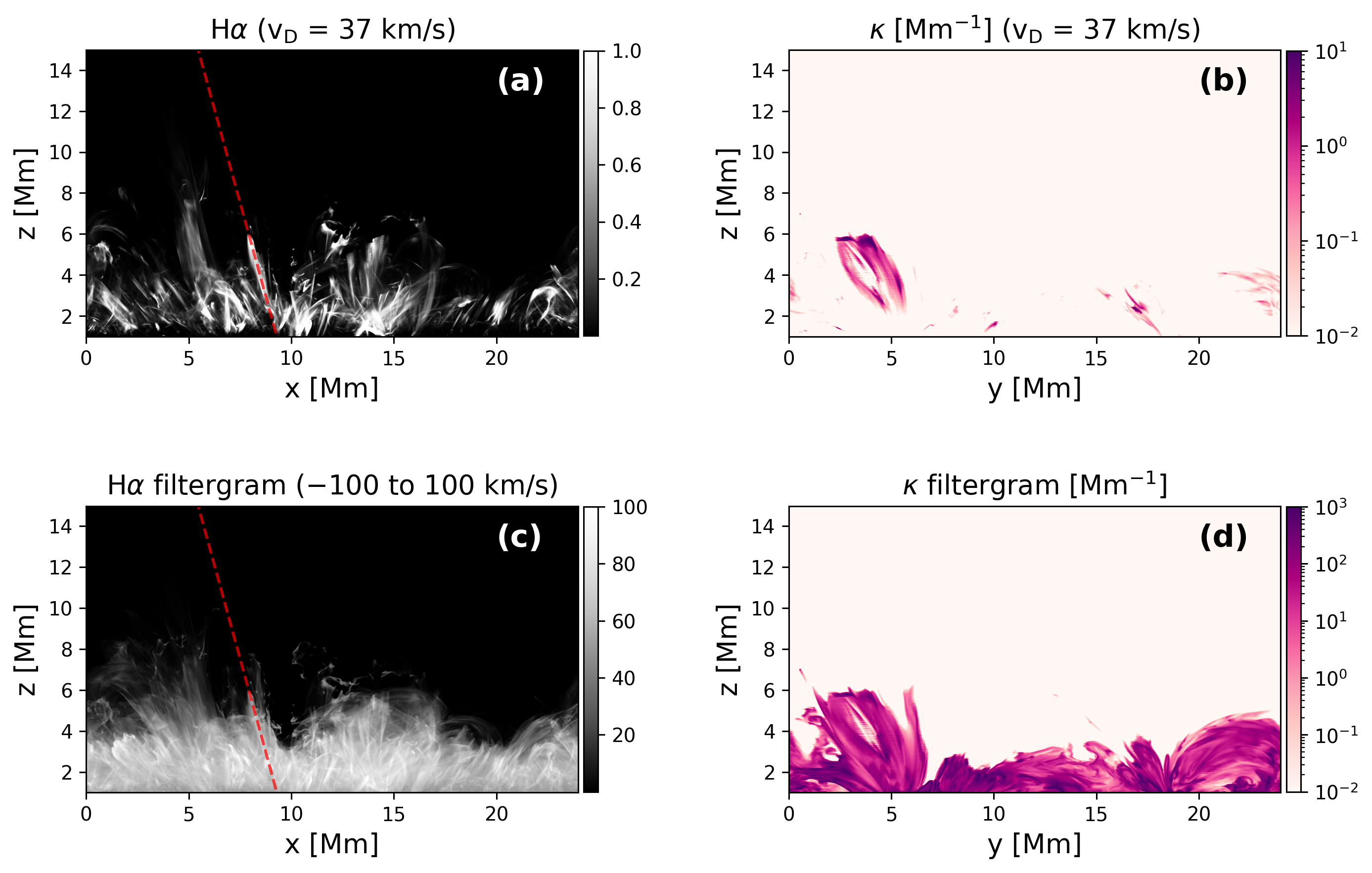}
   \caption{Spicule A shown in the H$\alpha$ filtergram. Panel~(a): Spicule A shown at a Doppler velocity of 37\;km/s. Panel~(b): The corresponding $\kappa$ slice at the same Doppler velocity. Panel~(c): The spicule in the H$\alpha$ proxy filtergram (integrated from $\mathsf{v}_\mathrm{D}$ = $-$100\;km/s to 100\;km/s). Panel~(d): The corresponding $\kappa$ filtergram, showing the slab-like structure. The dashed red line in panels (a) and (c) marks the slice through the spicule of interest.}
              \label{filtergram}%
    \end{figure*}


\section{Dynamics of representative spicules}\label{dynamics_spicules}

Features identified with the H$\alpha$ proxy are sensitive to the LoS velocities or the Doppler velocity ($\mathsf{v}_\mathrm{D}$). This is distinct from the apparent velocity of spicules observed in the plane-of-sky seen in the time-distance plots (e.g. panel b, Fig.~\ref{overview_F1}). A key distinction between the apparent velocities and the Doppler velocities is that the former may not be true mass motions, whereas the latter is. We adopt two Doppler velocities ($\mathsf{v}_\mathrm{D}$) of 37 km/s and 50 km/s which serve as our passbands to identify and investigate individual events. The choice of using $\mathsf{v}_\mathrm{D}$\;=\;50~km/s reduces contamination from overlapping structures (see \citealp[]{Chandra_2026}). But such high Doppler velocities sometimes may capture only part of the plasma composed of multiple velocity components. Thus combining this with a lower Doppler-velocity ($\mathsf{v}_\mathrm{D}$\;=\;37~km/s) analysis provides a more comprehensive picture. We analyse the temporal evolution of two spicules in detail. Examination of their structure using the 3D opacity ($\kappa$) information (see Sect. \ref{feature-of-interest}) reveals that one feature exhibits a slab-like morphology (spicule A), while the other displays a singly-threaded configuration (spicule B). These two cases are described in the following subsections.


    \begin{figure*}
   \centering
   \includegraphics[width=\textwidth]{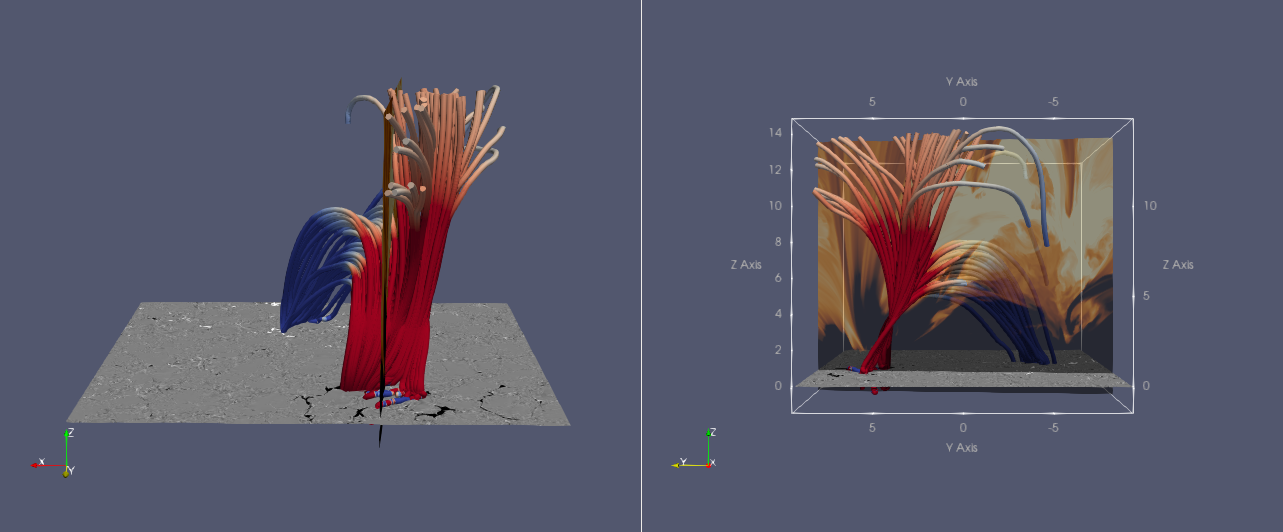}
   \caption{Magnetic topology around spicule A. Left: Side view of the topology across the slab. Right: Front view of the topology. The greyscale shows the vertical component of the magnetic field at the photosphere. The magnetic field lines are coloured by the vertical component of the magnetic field. The temperature slice highlights the location of the slab in the front view, and is plotted on both panels. The negative values on the y-axis in the right panel are computed using the periodic boundary condition in the horizontal direction. The slab is located at y\;$\approx$\;5\;Mm. An animation in time can be found \href{https://owncloud.gwdg.de/index.php/s/xuGKNtKDv5k7QvD}{online}.}\label{3D-sheet}%
    \end{figure*}


\subsection{Spicule A}

We analysed a spicule exhibiting a high apparent speed (in the plane of sky) of approximately 190~km/s during the initial $\sim$\;24~s of its evolution. This is followed by a slower secondary phase with a speed of about 25~km/s, lasting for an additional $\sim$\;42~s. The temporal evolution of this spicule is shown in panels (a) and (b) of Fig.~\ref{overview_F1}. This is the same spicule shown as an example slab-like spicule in Fig.~\ref{sheet-510000}. In panel (f) of Fig.~\ref{sheet-510000}, where we show an orthogonal view of the spicule, we also see a broader dim structure very close to the slab. We also identify an associated on-disc counterpart, which appears as an RBE, shown in panel (c) of Fig.~\ref{overview_F1}. The RBE synthesis assumes the vertical (z) as the line of sight, whereas the line of sight for the spicule is the horizontal (y). This initial analysis is done at a Doppler velocity of 37~km/s. The opacity contribution is further outlined in the $\kappa$ slice from the top view (panel d, Fig.~\ref{overview_F1}). The associated movie with Fig.~\ref{overview_F1} shows the evolution of the feature as a spicule and an RBE. It can be seen that the RBE is extended along the line of sight for the spicule (y axis), which is consistent with the slab picture. Moreover, the RBE exhibits a C-shaped trajectory as it evolves. This implies that the spicule moves in and out of a plane that passes through the spicule axis and is parallel to the line of sight (y axis).

The 3D structure of spicules plays a key role in interpreting their dynamics. The spicule examined here appears thin and sharply defined in the line-of-sight integrated maps (panel a, Fig.~\ref{overview_F1}). We then extract the opacity ($\kappa$) slice to estimate the 3D morphology. This reveals that a slab-like structure is oriented approximately along the viewing direction (panels b, c of Fig.~\ref{F1_various_vD}). The slab is visible throughout its lifetime at the lower Doppler velocity ($\mathsf{v}_\mathrm{D}$) of 37~km/s. At the higher Doppler velocity, the slab becomes less apparent during part of the evolution (panels d, f of Fig. \ref{F1_various_vD}). Thus the spicule also appears fainter at the Doppler passband of 50~km/s. The ratio of the contribution depth ($\mathrm{CD_{\tau = 1}}$) of the slab to the spicule width (FWHM) is $\approx$ 6.8. It can also be seen that at different stages of spicule evolution (e.g. panels A,B,C of Fig.~\ref{F1_various_vD}), either parts or the whole of the slab-like structure is visible based on the C-shaped trajectory in the on-disc view. We also synthesised a H$\alpha$ proxy filtergram covering Doppler velocities from $-$100\;km/s to 100\;km/s and found that the slab-like structure remains visible throughout the lifetime of the spicule. However, there are substructures in the slab on all length scales (see Fig.~\ref{filtergram}). We note that this broad passband is very similar to that of the Ca II H filtergraph on \textit{Hinode}/SOT used for observing spicules, which is about 0.3\;nm or $\pm$113\;km/s.

We investigated the magnetic connectivity, wherein we tracked the temporal evolution of the feature and traced a 3D magnetic field line threading it, following the approach of \citet{Chandra_2025}. During the evolution, this field line transitions from a closed to an open configuration, and subsequently returns to a closed state \footnote{Open or closed field lines only indicate whether magnetic topology is closed with both footpoints connecting to the solar surface, or open with only one footpoint connecting to the surface within the simulation domain.} The magnetic topology surrounding the slab is shown in Fig.~\ref{3D-sheet}. The local magnetic topology is characterised by open field lines rooted in the strongest network concentrations and by large loops connecting to more distant opposite-polarity regions. To indicate the slab’s position, we use a temperature slice in the plane of the slab, where the chromosphere is displayed in dark brown. Field lines traced across this plane reveal that the slab lies at the interface between open and closed field lines. In other words, the magnetic topology changes across the slab, indicating that it forms within a quasi-separatrix layer (QSL), where magnetic field lines transition sharply between open and closed configurations. A detailed analysis with the squashing factor is shown in Appendix~\ref{QSL_appendix}.


   \begin{figure*}
   \centering
   \includegraphics[width=\textwidth]{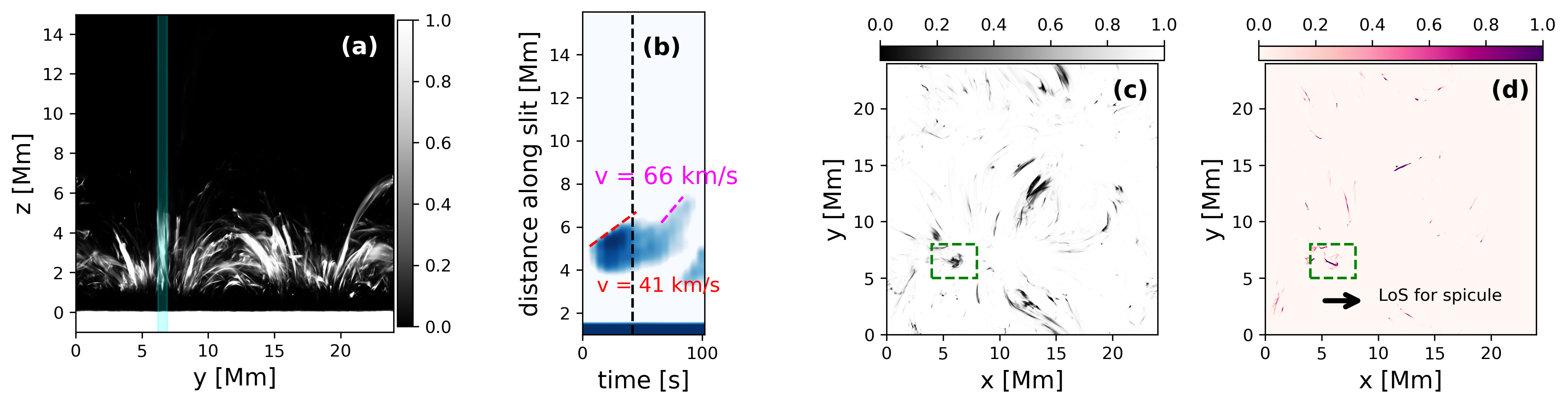}
   \caption{Overview of the evolution of singly-threaded spicule B. The panels represent the same quantities as in Fig. \ref{overview_F1}. The opacity slice in panel (d) is shown at a constant height of 4.3\;Mm. A movie can be found \href{https://owncloud.gwdg.de/index.php/s/OOgMak6EewC79Cv}{online}.}
              \label{overview_F2}%
    \end{figure*}

   \begin{figure*}
   \centering
   \includegraphics[width=\textwidth]{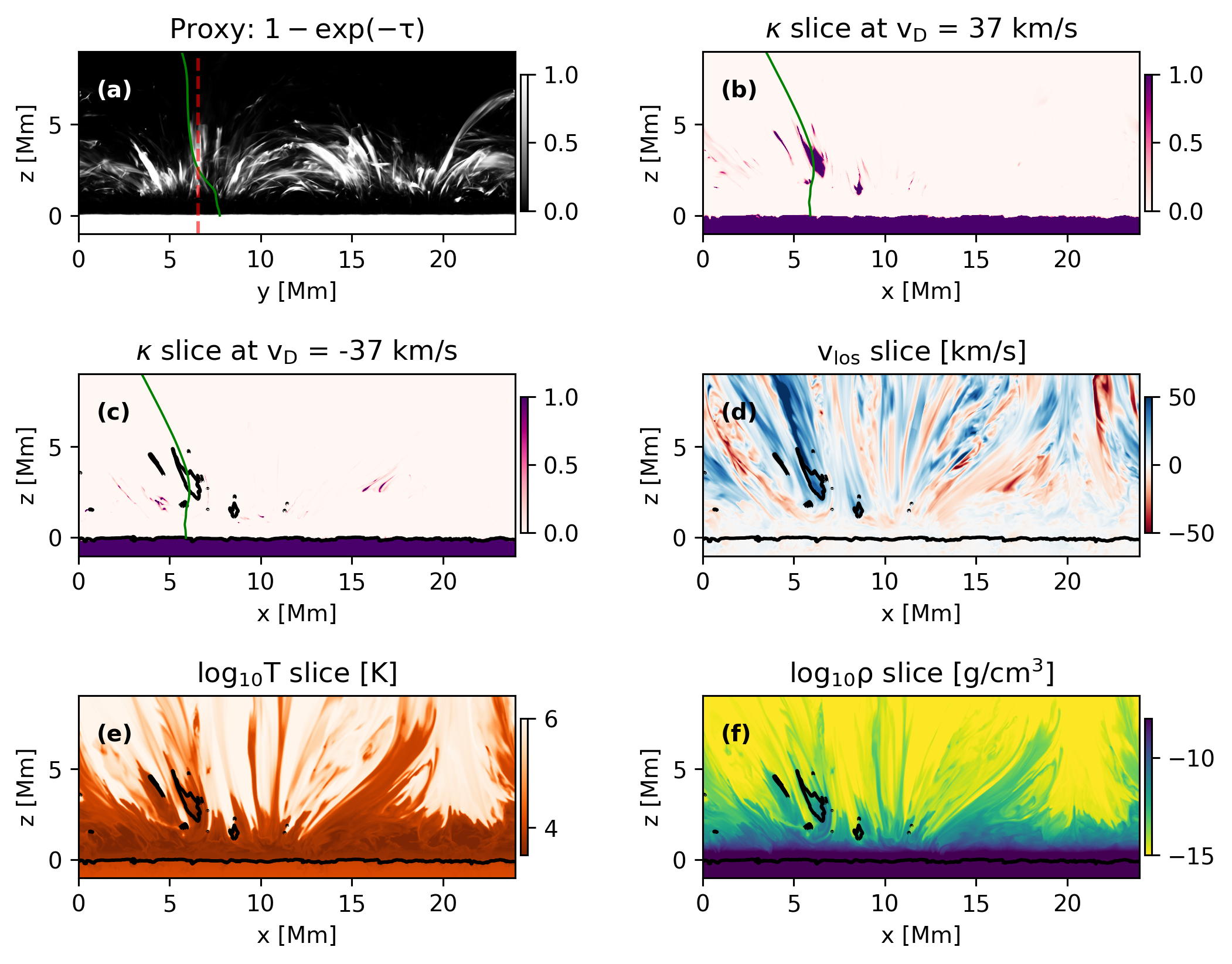}
   \caption{Swaying of spicule B in the line of sight. Panel (a) shows the feature in the proxy with the location of the slice marked by the dashed red line. Panels (b)-(f) show the same slice in the opacity at 37~km/s, opacity at $-$37~km/s, LoS velocity, temperature, and density. The field line threading the feature is plotted as the green line in panels (a), (b), and (c). The black contour lines outline the feature in the opacity seen in panel (b). A movie showing the evolution of the spicule from the blue wing to the red wing can be found \href{https://owncloud.gwdg.de/index.php/s/VG60651ZkMcsbjL}{online}.}
              \label{s2_wobbling}%
    \end{figure*}

    \begin{figure}
   \includegraphics[width=0.5\textwidth]{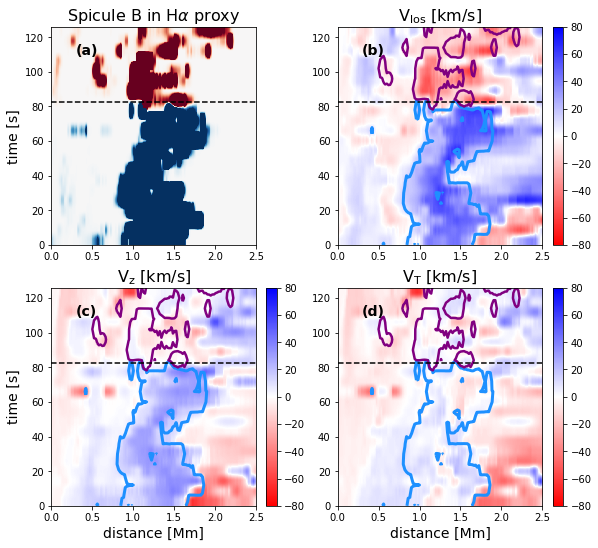}
   \caption{Time-distance plots showing the evolution of spicule B and the various components of the velocity. Panel (a) shows the evolution of the feature in the proxy at $\mathsf{v}_\mathrm{D}$ = $\pm$37~km/s. The dashed horizontal black line shows the time when the feature transforms from being blueshifted to redshifted. Panels (b) - (d) show the LoS velocity, vertical velocity component ($\mathsf{v}_\mathrm{z}$), and flows projected along the tangent to the field line threading the feature ($\mathsf{v}_\mathrm{T}$). Blue and purple contour lines mark the region occupied by the feature, as seen in the blue and red wings, respectively. }
              \label{time_distance_plot}%
    \end{figure}

    \begin{figure}
   \centering
   \includegraphics[width=0.5\textwidth]{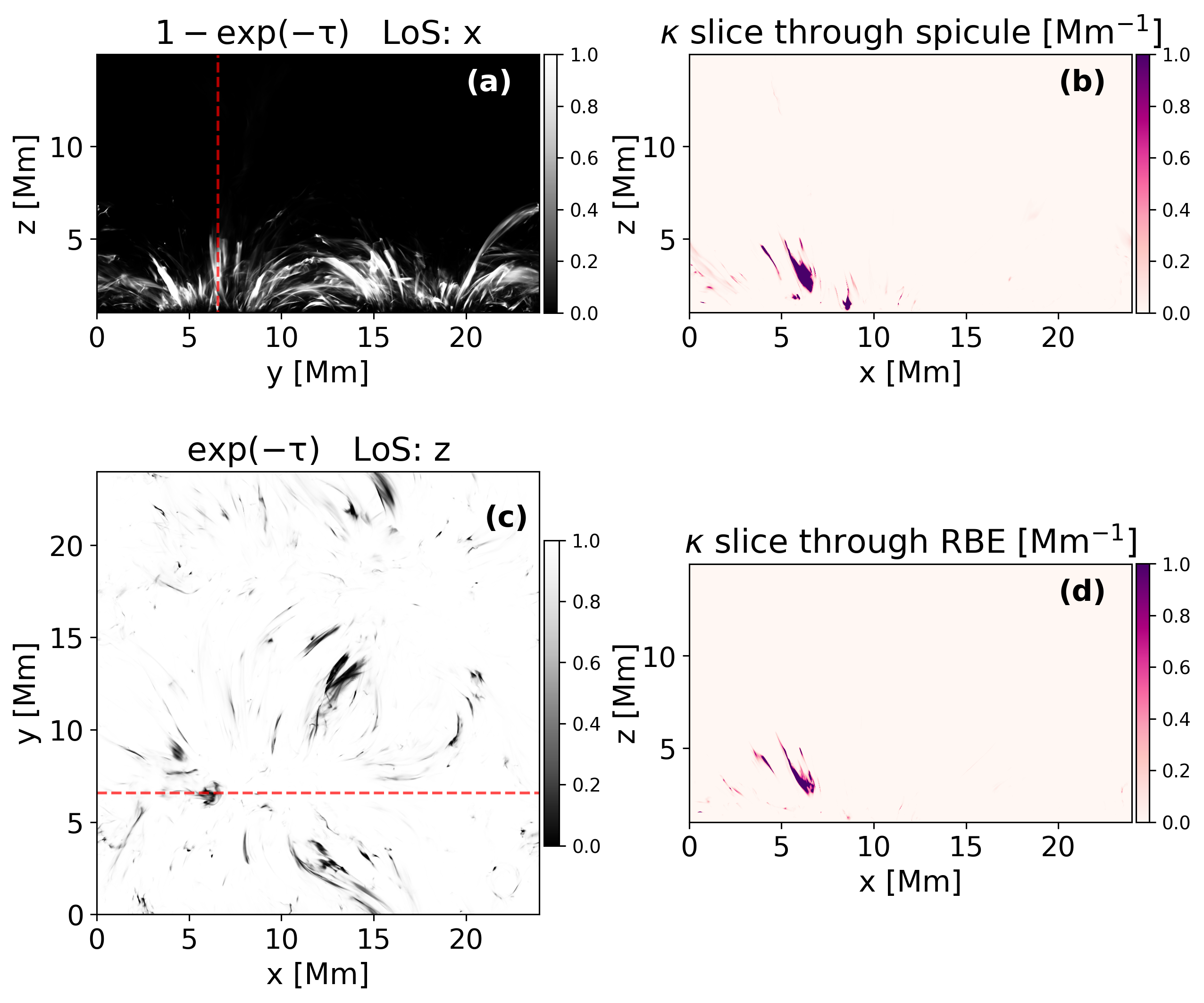}
   \caption{Opacity along the line of sight for the off-limb spicule (spicule B) and the on-disc RBE synthesised at $\mathsf{v}_\mathrm{D}$ = 37\;km/s. Note that the 37\;km/s is in the z direction in the case of the RBE analysis and in the x direction for the off-limb spicule. Panel (a) shows the spicule and the slice marked by the dashed red line, while panel (b) shows the contribution from the opacity along the line of sight (x) in the slice shown in panel (a). Panel (c) shows the RBE feature associated with the spicule with the dashed red line marking the slice, while panel (d) shows the contribution from the proxy in this slice, in the line of sight (z). The evolution of the feature from the two viewing angles can be found \href{https://owncloud.gwdg.de/index.php/s/HR29CaV3Zm9Gt3U}{online}.}
              \label{s2_contribution}%
    \end{figure}


Thus, spicule A exhibits a slab-like morphology, with a clear RBE counterpart in the synthesised on-disc view. Our analysis also demonstrates that the slab is formed at the interface of open and closed field lines, in a QSL. As the field evolves, the movement of the plasma sometimes aligns with the line of sight, rendering a sharp appearance of the spicule moving rapidly upwards. These results indicate that the sharp apparent upward motion (of $\sim$ 190\;km/s) cannot be explained simply by plasma flowing along a static open field line but is instead consistent with the projection effect of a dynamically evolving slab-like structure. Furthermore, the average vertical upflow in the spicule reaches a maximum of only 39~km/s.

\subsection{Spicule B}

We examined a second spicule at a Doppler velocity ($\mathsf{v}_\mathrm{D}$) of 37~km/s, which exhibits a maximum apparent upward velocity of approximately 66~km/s. Its evolution is more complex than that of spicule~A, showing an initial upward motion, a partial fallback, and a subsequent re-acceleration. The temporal evolution of this feature is illustrated in Fig.~\ref{overview_F2}. An on-disc synthesis revealed an RBE counterpart to this spicule, analogous to the case of spicule~A. This RBE is a broader structure that gradually transforms into a sharp feature, as viewed on disc (see the movie associated with Fig.~\ref{overview_F2}).

We highlight the singly-threaded structure in the LoS contribution to the opacity (panel~b of Fig.~\ref{s2_wobbling}). This is also evident in the temperature and density profiles (panels~e, f of Fig.~\ref{s2_wobbling}). The vertical velocity slice (panel~d of Fig.~\ref{s2_wobbling}) shows upflows consistent with those associated with the RBE at the same location.

The lifetime of this feature is about 80 seconds, after which it quickly disappears from the $\mathsf{v}_\mathrm{D}$ = 37\;km/s bandpass. To understand why it disappears, we examined a slice through the feature along the LoS velocity ($\mathsf{v}_\mathrm{x}$ in this case). The velocity rapidly transitions from motion towards to motion away from the observer (see panel~d of Fig.~\ref{s2_wobbling} and the accompanying movie). The redshifted flows reach speeds of 35~km/s–50~km/s. To explore this behaviour further, we synthesised proxy slices at both +37~km/s (towards the observer) and –37~km/s (away from the observer). This revealed a clear continuation of the feature as it evolves from the blue to the red wing of the H$\alpha$ line (panels~b, c of Fig.~\ref{s2_wobbling}). Thus spicule B appears to move back and forth along the line of sight, explaining its apparent disappearance in the original bandpass of 37\;km/s. This highlights the role of complex motions in shaping the dynamics of spicule-like features, consistent with their observational signatures \citep[e.g.][]{Sekse_2013b}.

We also tracked spicule~B throughout its lifetime and traced a magnetic field line threading it. In this case, the field line connectivity remained open throughout the evolution. Using these traced field lines, we generated time–distance plots by stacking them in time, following the method of \citet{Chandra_2025}. Figure~\ref{time_distance_plot} shows the evolution in the time-distance plot, where the feature transitions from the blue to the red wing of the H$\alpha$ spectral line. The vertical velocity component ($\mathsf{v}_\mathrm{z}$) extracted along the traced field lines reveals that the blue-wing phase corresponds to upflows, consistent with the RBE counterpart identified for this spicule. Furthermore, the LoS velocity component ($\mathsf{v}_\mathrm{los}$) shows a clear transition from blueshifts to redshifts within the region occupied by the feature (highlighted with contour lines). This suggests that spicule lifetimes may often be underestimated when measured in a single Doppler velocity channel, as they can shift between different LoS velocities during their evolution. Such complex motions have been reported in spicules observed in the H$\alpha$ line \citep{Pereira_2016}.

To determine which type of flow -- transverse or field-aligned -- dominates the dynamics of this spicule, we projected the velocity field onto the tangential (T) direction, aligned with the traced magnetic field line. The resulting field-aligned flows ($\mathsf{v}_\mathrm{T}$) are very weak ($\sim$\,10~km/s) compared to the LoS velocities at $\sim$\,30~km/s–50~km/s. This indicates that field-aligned motion is small and the observed feature moves faster along the line of sight, which is at 90 degrees with the field line itself.

Spicule~B appears to have a more singly-threaded 3D structure, although its appearance varies with Doppler velocity. Towards the end of its evolution in the blue wing, it becomes more extended and resembles a slab. We also investigated the magnetic topology around spicule B, which revealed its location at a QSL. The RBE associated with spicule~B initially appears as a broader structure that later evolves into a sharper, extended feature on disc (see the movie associated with Fig.~\ref{overview_F2}). The disappearance of spicule~B from the original bandpass is explained by transverse motions along the line of sight. Throughout its evolution, this transverse motion dominates over the field-aligned flows.

\section{Discussion}\label{discussion}

Using our H$\alpha$ proxy, we have identified and analysed 58 spicule-like structures that closely resemble their observed counterparts \citep{Chandra_2026}. With MURaM-ChE, we report on the first type~II spicule formations in 3D, revealing a diverse range of morphologies -- from narrow, thread-like features to broader, slab-like structures. Some of the structures also appear as multi-threaded spicules (see Appendix~\ref{morphology_appendix}, Fig.~\ref{multithread}).
Their evolution is highly dynamic and complex, consistent with theoretical expectations that some spicules may result from optical illusions \citep{Judge_2011}, as well as with high-resolution observations \citep{Pereira_2012, Sekse_2013b, Pereira_2016}. The very high apparent upward motions (for spicule A) are not real mass flows. Instead, they arise from the motions of the slab-like structure aligning with the line of sight.

Previous numerical investigations have studied spicules using various approaches. For example, \citet{Sykora_2017, Sykora_2020} demonstrated in 2.5D simulations that ambipolar diffusion can generate fast and tall spicule-like jets which appear as conical protrusions in density maps. More recently, \citet{dey_2024} reported fully 3D simulations that produced spinning spicular structures without imposed flux emergence, attributing their formation purely to the pleating of a sheet-like plasma structure, or “drapery”.

Although observations consistently show that spicules cluster around network magnetic fields, the detailed magnetic topology of their environment remains poorly constrained. In our earlier work \citep{Chandra_2025}, we characterised the magnetic structure associated with an RBE and demonstrated field-line rearrangement following flux cancellation. Building on this, we now examine the magnetic configuration linked to the taller spicules (reaching >3\;Mm above the solar surface), providing new insight into how these features connect to their magnetic environment.

Finally, we investigated the kinematics of the singly-threaded spicule B. We find that its apparent motion is dominated not by field-aligned flows but by strong transverse displacements along the line of sight, challenging the traditional picture of plasma being passively guided along quasi-static magnetic field lines. We also briefly discuss the on-disc counterparts of the two off-limb spicules analysed in this study. In the following, we explore these aspects in more detail.

\subsection{Morphology and motion of spicules}\label{morphology_motion}

The nature of spicules has long been debated. Early studies proposed that they are one-dimensional expanding magnetic flux tubes extending from the photosphere into the corona \citep{Sterling_1993, Kudoh_1999}. In the 2.5D simulations of \citet{Sykora_2017, Sykora_2020}, spicules were identified as conical protrusions in temperature and density maps based on the 2D geometry. An alternative interpretation suggests that spicules arise from ripples in two-dimensional sheets viewed edge-on, giving the illusion of narrow, conical spikes \citep{Judge_2011, Judge_2012}. Our simulations reveal the presence of a diverse range of morphologies, emphasising that spicules are intrinsically complex and cannot be adequately described by simple geometric models. In particular, the slab-like spicules we identify differ fundamentally from structures produced by ripples in a sheet, in that the entire slab contributes to spicule visibility unlike only ripples in a larger structure. While multi-threaded spicules may, in some cases, be consistent with ripples in a larger sheet, such configurations represent only a small fraction of the spicules found in our simulation.

Observationally, the rapid apparent motions and short lifetimes of spicules—along with their disappearance from chromospheric passbands and reappearance in hotter channels—have often led to the interpretation that they are jet-like structures supplying mass and energy to the upper solar atmosphere. However, such interpretations remain uncertain without a full 3D picture of their motion. Most early models implicitly assume that spicules follow field-aligned flows within cylindrical flux tubes, producing apparent upward and downward motions (see e.g. the review by \citealp[]{Sterling_2000}). More recent studies show that the dynamics of spicules are governed by the interplay of complex motions \citep{Sekse_2013a, Shetye_2016}. In our case we find that for spicule B, cross-field motions dominate its dynamics. A similar inference can be found in \citet{Chandra_2025}, where the field-aligned flows in an RBE were much weaker than the vertical (or LoS) motions. To illustrate the complex behaviour of the simulated spicules, we provide two examples.

Spicule A exhibits a slab-like morphology aligned with the line of sight, producing a sharply defined feature in the H$\alpha$ proxy at the limb. Its on-disc RBE counterpart also appears as an extended structure (panel c, Fig.~\ref{overview_F1}). The feature shows an apparent upward speed of about 190~km/s, which does not correspond to real mass motions but arises from the motion of the 3D slab. We further computed the average vertically upward flow in the location of the slab, which reaches a maximum of 39\;km/s. The associated RBE follows a C-shaped trajectory toward the magnetic network patch, showing abrupt changes in apparent speed as its direction shifts. This change is also seen in the apparent velocity of the spicule itself (see panel b of Fig.~\ref{overview_F1}). When viewed at the limb, the resulting projection may give the illusion of a swirling motion, even though no true rotation is present. This case highlights the complex multi-component dynamics of spicules in 3D and the challenges inherent in interpreting their observations using only 2D projections.

Spicule B displays a singly-threaded morphology through a part of its evolution. Its RBE counterpart initially appears broad and diffuse but becomes increasingly elongated and slab-like toward the end of its lifetime (see the movie associated with Fig.~\ref{overview_F2}). This evolution illustrates that the morphology of spicules can change significantly over time. Furthermore, the morphology is sensitive to the Doppler passband used for the analysis. 

The motion of spicule B shows that the magnetic field line itself moves rapidly along the line of sight, with only very weak field-aligned plasma flows (Fig.~\ref{time_distance_plot}). At the same time, the feature exhibits upward motion due to vertical plasma flows (panel c, Fig.~\ref{time_distance_plot}), thereby showing an RBE counterpart. Its LoS velocity alternates between blue and red-shifts, indicating swaying motions. This implies that spicules that seem to vanish may simply reappear at a different Doppler velocity, exhibiting rapid disappearance and short lifetimes when observed at a single Doppler passband, similar to the results of \citet{Pereira_2016}.

The evolution of the 3D structure of spicules is therefore highly complex, and their on-disc counterparts can differ in appearance from the off-limb structures. This raises the question of whether RBEs and spicules indeed share the same physical structure, an issue explored further in Sect.~\ref{RBEs_as_spicules}.

\subsection{RBEs as counterparts of type II spicules}\label{RBEs_as_spicules}

For both spicules A and B, we identified corresponding RBE counterparts. Previous observations \citep{Rouppe_van_der_Voort_2009, Sekse_2012, Sekse_2013b} have shown that type II spicules and RBEs/RREs share similar characteristics, including lifetimes, apparent speeds, widths, and both transverse and torsional motions. However, observations alone provide limited insight into their physical connection.

Our simulations reveal that the RBEs associated with the two type II spicules exhibit very similar opacity structures along the line of sight. An example for the spicule B is shown in Fig.~\ref{s2_contribution} and the associated movie. The opacity signature for spicule B, synthesised from the horizontal direction and its RBE counterpart synthesised from the vertical direction, appear very similar throughout its lifetime. We find a comparable result for spicule A, although the appearance of the opacity structures differ slightly owing to the inclination of spicule A with respect to the vertical. This agreement confirms that the spicules we studied indeed have associated on-disc RBE signatures. Because the proxy synthesis is performed from different viewing angles, we can examine the opacity contribution at independent LoS velocities. That the opacity contribution appears very similar confirms that we are observing the same feature.

We also find that the apparent motions of the RBEs (projected on the solar disk) differ from those of the spicules observed off-limb. Taken together, these results provide a comprehensive picture of the dynamics of small-scale structures as seen in the H$\alpha$ spectral line.

\subsection{Magnetic topology associated with the spicules}\label{magnetic_topology}

We investigated the magnetic topology across all the spicules extending beyond 3\;Mm above the mean solar surface. All these spicules are associated with QSLs in our simulation. We show an example with spicule A, where we find a spatial transition in the field-line connectivity, from lines that are fully anchored in the photosphere (closed field lines) to those with only one magnetic footpoint rooted in the photosphere (open field lines). This transition defines a quasi-separatrix layer (QSL), with the slab located at the interface where the magnetic topology changes (see Figs.~\ref{3D-sheet} and \ref{QSL}). A similar topological analysis for spicule B shows that it is also associated with a QSL. 

\section{Conclusion}\label{conclusion}

The morphology of spicules has long been debated, with a thread-like structure traditionally being the dominant interpretation. Our results highlight the need for a fully 3D view of spicules -- something that remains beyond the capabilities of current remote-sensing observations. Using the enhanced-network MURaM-ChE simulation, we demonstrate that highly dynamic spicules with both slab-like and thread-like morphologies coexist. The high apparent upward speeds observed in some spicules may originate from transverse motions of a larger, slab-like structure. These high apparent speeds do not represent real mass (up)flows, which are found to be much smaller in magnitude. 

We stress that we exclude ambipolar diffusion in our simulation. Nevertheless, we observe self-consistently formed spicules, including type II spicules, similar to their observed counterparts. This suggests that ambipolar diffusion is not a necessary ingredient for the formation of spicules. This finding is in disagreement with the work of \cite{Sykora_2017}, which claims that the formation and subsequent evolution of the taller and more dynamic spicules depend on the process of ambipolar diffusion.

The magnetic topology in the vicinity of the spicules in this study were found to be QSLs, i.e. most of the spicules are located at the interface between open and closed field lines. The magnetic field at higher heights (3\;Mm above the solar surface) support the plasma structure appearing either as slabs or as threads. However, what gets the plasma up to these heights remains a matter of debate. This is likely related to magnetic driving at the base of the spicules and shock fronts propagating from the base of the chromosphere towards the corona.

Spicules also display highly dynamic behaviour, including pronounced transverse motions along the line of sight. In the analysis of spicule B, transverse motions were found to significantly contribute to its dynamics. This spicule evolves from a blueshifted to a redshifted profile, indicating that Doppler observations at a single wavelength may systematically underestimate spicule lifetimes and evolution \citep{Pereira_2016}. This further implies that the disappearance of type II spicules involves processes beyond simple heating to higher temperatures.

We identified on-disc counterparts of spicules that exhibit consistent opacity signatures across spectral slices at orthogonal line-of-sight velocities, that is, when observed on the disc and off the limb. This provides a direct confirmation that spicules and RBEs represent the same underlying chromospheric structures in H$\alpha$, when analysed from two orthogonal vantage points. Such an analysis is only possible using 3D simulations of spicules.

We use narrowband synthetic images at a fixed Doppler velocity for most of our analysis. However, the morphology is dependent on the Doppler velocity and the evolution phase of the spicule. When comparing with broadband filtegrams of the spicules, we find that the morphology is largely preserved at the Doppler velocity of $\pm$\;37\;km/s. An example is shown in Fig.~\ref{filtergram}. For the few multi-threaded spicules we identify, we also find the same structure in the H$\alpha$ filtergram image. These could sometimes be a part of a larger structure, similar to folds in a sheet \citep{Judge_2011}. An analysis through time, could give hints into the group behaviour of spicules, connected with the multi-threaded spicules, but is beyond the scope of this paper.

In summary, spicules are inherently three-dimensional and dynamically complex. Interpreting their nature requires realistic 3D simulations with synthetic observables that bridge the gap between models and current observational constraints. However, because we purely use a H$\alpha$ proxy for the present study, we miss information on what happens to the spicules after they disappear from the H$\alpha$ passband altogether. To check whether some of these features are heated to transition region temperatures (e.g. in Si~IV) and are further visible in hotter coronal channels (\citealp[e.g.][]{Bose_2025}) in the simulation, is a matter of future investigation. Some of the very high speeds associated with spicules have also been attributed to heating fronts in previous studies \citep{de_Pontieu_2017}. Our previous work \citep{Chandra_2025} also shows a heating front associated with the RBE from the same simulation. To investigate the same phenomenon for spicules is beyond the scope of this study.

\begin{acknowledgements}
       We thank T. M. D. Pereira for useful comments which greatly improved the paper. S.C. thanks P. Dey for useful discussions. This work was carried out in the framework of the International Max Planck Research School (IMPRS) at the Technical University of Braunschweig. We are grateful for the computational resources provided by the Raven and Viper supercomputer systems of the Max Planck Computing and Data Facility (MPCDF) in Garching, Germany. This research has received financial support from the European Research Council (ERC) No. 101097844 (WINSUN). This work was supported by the Deutsches Zentrum f{\"u}r Luft und Raumfahrt (DLR; German Aerospace Center) by grant DLR-FKZ 50OU2201. 
\end{acknowledgements}

\bibliographystyle{aa} 
\bibliography{bibfile} 

\onecolumn
\appendix

\section{Spicules with different morphologies}\label{morphology_appendix}

\begin{figure}[ht!]
   \centering
   \includegraphics[width=\textwidth]{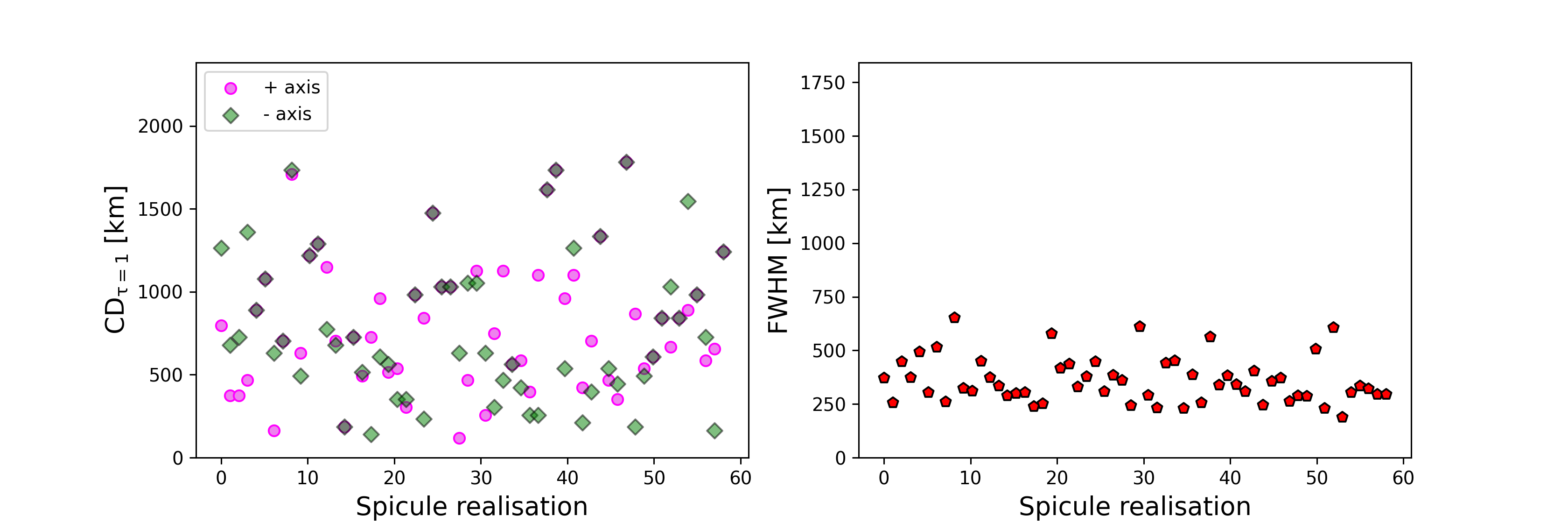}
   \caption{Distribution of contribution depths ($\mathrm{CD_{\tau = 1}}$) and spicule widths (FWHM) over all 58 spicule realisations. Positive (+) and negative ($-$) axes on the left panel indicate the viewing direction along the positive and negative axes respectively.}
              \label{distribution}%
    \end{figure}

\begin{figure}[ht!]
   \centering
   \includegraphics[width=0.95\textwidth]{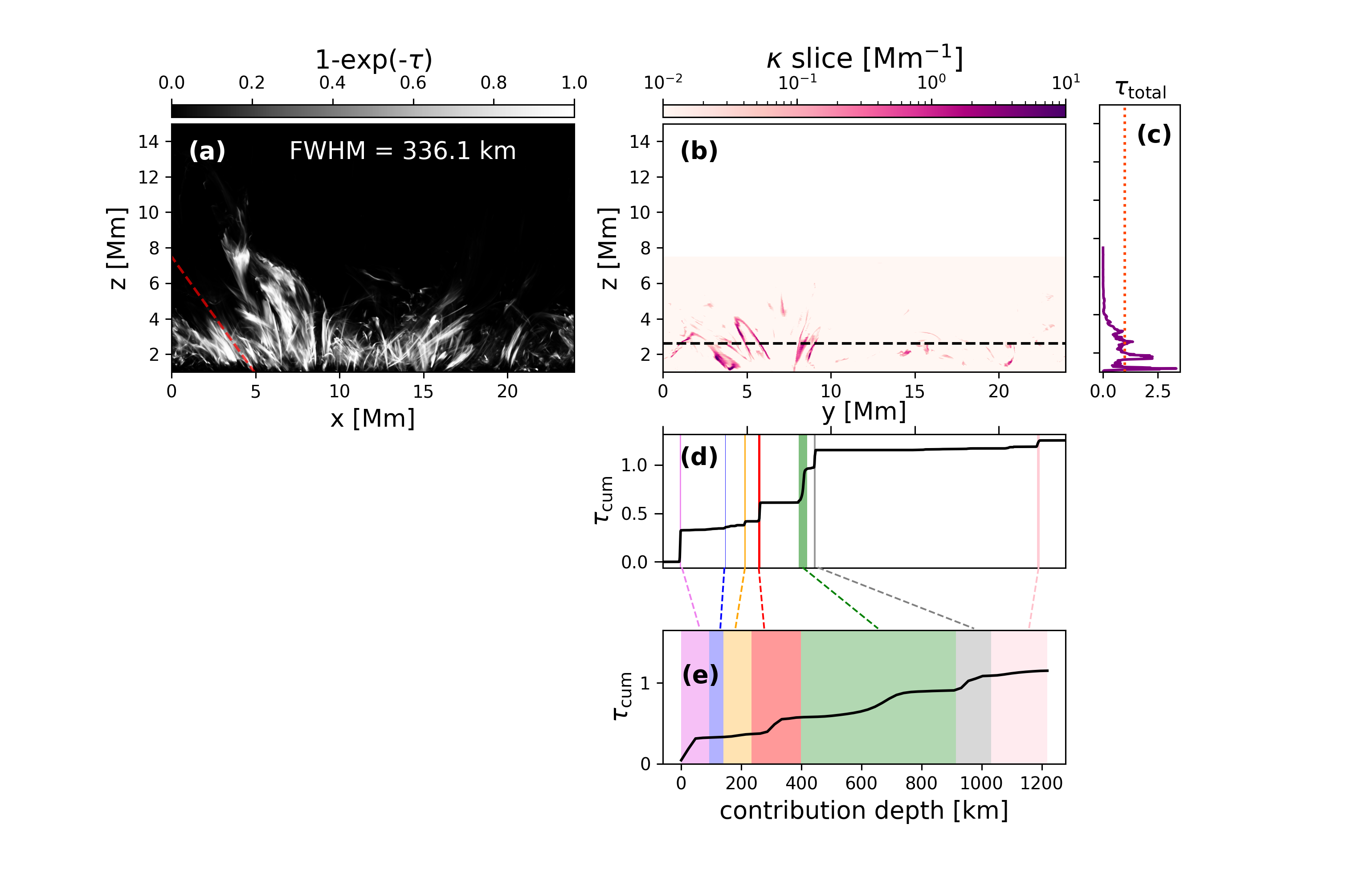}
   \caption{Overview of a multi-threaded spicule in a similar format as Fig~\ref{thread-535000}. The different regions contributing to the spicule visibility have been shaded in panels (d) and (e).}
              \label{multithread}%
    \end{figure}

\begin{figure}
   \centering
   \includegraphics[width=0.95\textwidth]{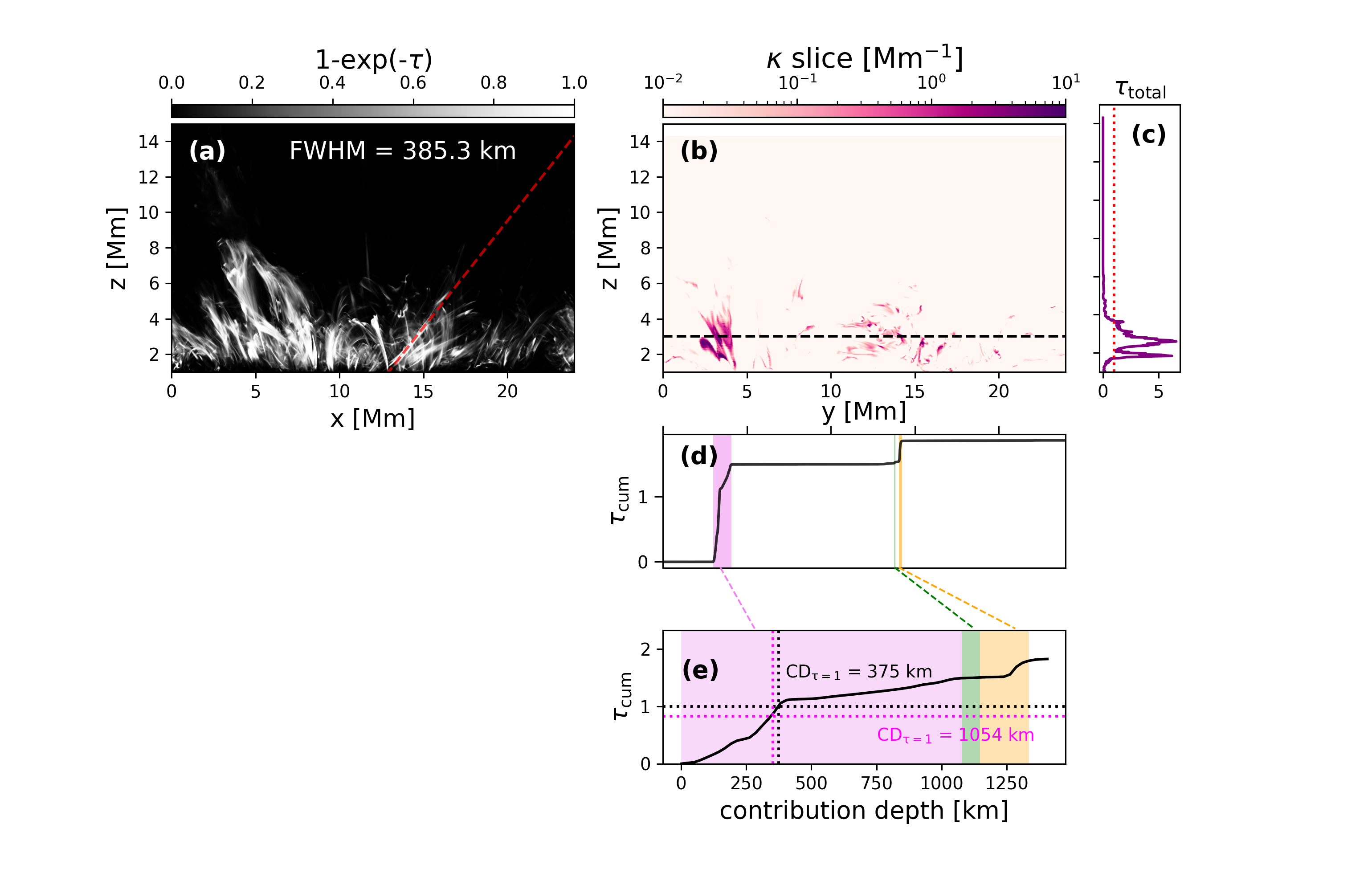}
   \caption{Overview of a spicule, appearing as a slab from one direction and as a thread from the opposite direction along the line of sight in the same format as Fig.~\ref{multithread}. The black dotted vertical and horizontal lines on panel (d) show the contribution depth as seen from the positive axis (as a thread), while the corresponding magenta lines show it from the negative axis (as a slab).}
              \label{sheet-thread}%
    \end{figure}

We present four categories of spicule morphologies in Table~\ref{table:1}. The classification into thread-like or slab-like is based on the ratio of contribution depth to spicule width, noting that the contribution depth depends on the viewing direction (along or against the line of sight). The distributions of contribution depths and widths for all 58 spicules in the enhanced network simulation are shown in Fig.~\ref{distribution}. We describe a slab-like spicule A and a singly-threaded spicule B in the main text. If visibility arises from spatially separated narrow opacity structures, the spicule appears multi-threaded; despite large contribution depth, the spatial separation distinguishes it from slab-like cases (see Fig.~\ref{multithread}). We also find cases that appear slab-like or thread-like depending on the viewing direction: along or against the chosen the line of sight (Fig.~\ref{sheet-thread}).

\newpage

\section{Computing the squashing factor for determining QSLs}\label{QSL_appendix}

\begin{figure}[ht!]
   \centering
   \includegraphics[width=\textwidth]{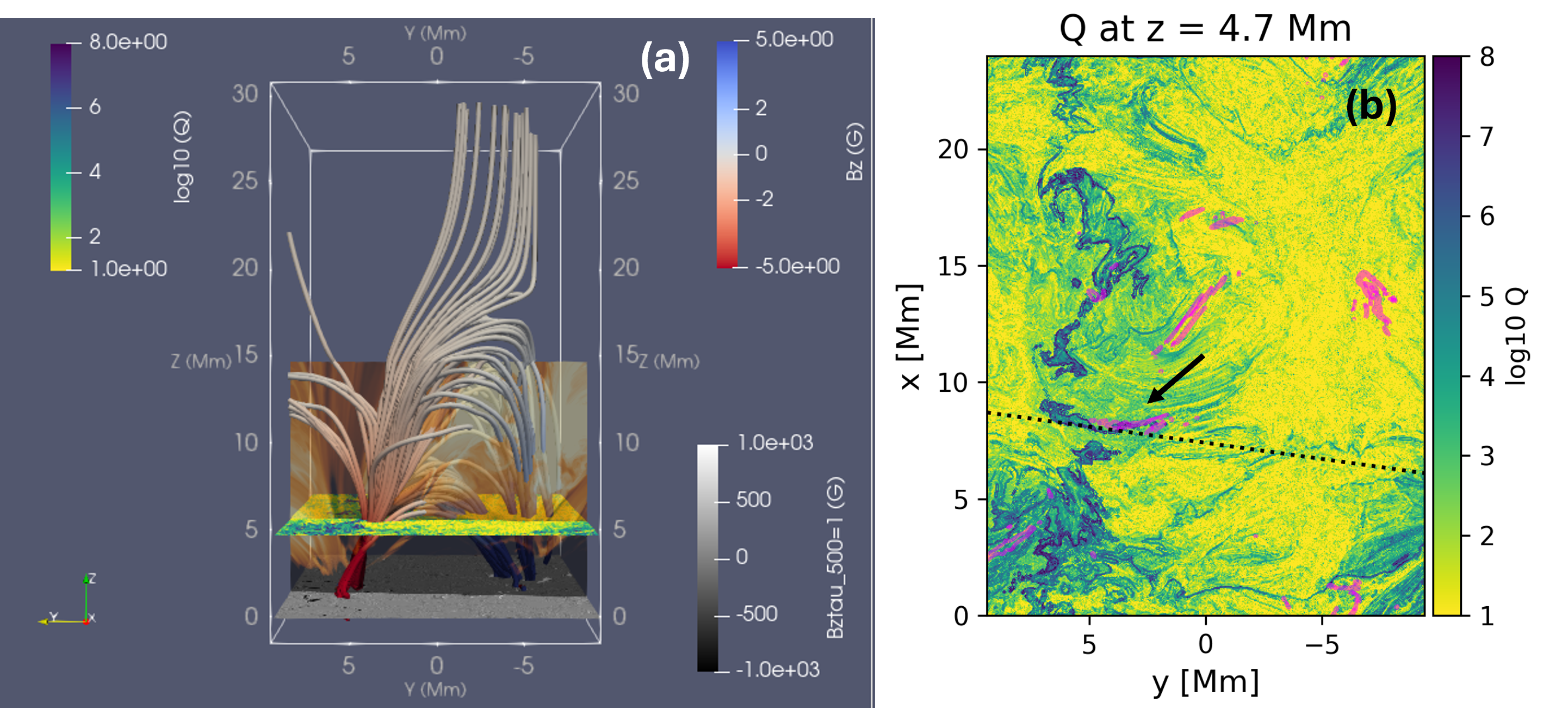}
   \caption{Squashing factor (Q) computed at a fixed height of 4.7\;Mm for spicule A. In panel (a), we show the extrapolated magnetic fields for an extended box with the top boundary at 30.6\;Mm above the solar surface. This is the same view as shown on the right panel in Fig.~\ref{3D-sheet}. In panel (b) we show the Q map at the chosen surface, with the temperature slice shown as a dotted black line cutting across the Q map. We also show the H$\alpha$ proxy ($\mathsf{v}_\mathrm{D}$\;=\;37\;km/s) at this height with the magenta contours, with the black arrow indicating the region of spicule~A.} 
              \label{QSL}%
    \end{figure}

The squashing factor (Q) quantifies the spatial gradients in the connectivity of magnetic field lines. Regions of high Q indicate strong gradients in field-line connectivity and are commonly associated with magnetic reconnection and quasi-separatrix layers (QSLs). Our original simulation box has a potential field top boundary condition for the magnetic field. To compute Q, we performed a potential field extrapolation, extending the vertical domain from 14.6\;Mm to 30.6\;Mm above the solar surface. This extension accounts for the proper distinction between open and closed field lines. The squashing factor was then evaluated on a horizontal slice at z\;=\;4.7\;Mm, as shown in Fig.~\ref{QSL}. The definition of Q follows Eq.~3 of \citep{Zhang_2022}, modified to include the connection height (z) of field lines, which is taken to be either the top boundary (z\;=\;30.6\;Mm) or the surface (z\;=\;0\;Mm). The resulting Q values reach up to $\sim 10^8$ in the region associated with spicule A (panel b, Fig.~\ref{QSL}). Not all opacity structures visible at a particular height give rise to spicules: The other opacity structures seen in panel (b) of Fig.~\ref{QSL} do not correspond to any of the 58 spicules that we identified in the simulation.

\end{document}